\documentclass[natbib,epjST]{svjour}
\usepackage{graphics}
\usepackage[hyphens]{url}
\usepackage[hyperindex,breaklinks]{hyperref}

\usepackage[sort&compress,numbers]{natbib}
\usepackage{doi}

\hypersetup{
  colorlinks=true,
  linkcolor=blue,
  filecolor=magenta,
  urlcolor=blue,
  citecolor=blue,
}

\begin{document}
\title{Non-perturbative methods for $NN$ singular interactions}
\author{D.R. Entem\inst{1}\fnmsep\thanks{\email{entem@usal.es}} \and 
J.A. Oller\inst{2}\fnmsep\thanks{\email{oller@um.es}}}
\institute{Grupo de F\'\i sica Nuclear e IUFFyM, Universidad de Salamanca, E-37008 Salamanca, Spain \and 
Departamento de F\'\i sica, Universidad de Murcia, E-30071 Murcia, Spain
}
\abstract{
Chiral Effective Field Theory ($\chi$EFT) has been extensively used to study the $NN$ interaction during the
last three decades. In Effective Field Theories (EFTs) the renormalization is performed order by order including
the necessary counter terms. Due to the strong character of the $NN$ interaction a non-perturbative resummation 
is needed. 
In this work we will
review some of the methods proposed to completely remove cutoff dependencies. 
The methods covered are renormalization with boundary conditions, renormalization with
one counter term in momentum space (or equivalently substractive renormalization) 
and the exact $N/D$ method.
The equivalence between the methods up to one renormalization condition will be checked showing results in the
$NN$ system. The exact $N/D$ method allows to go beyond the others,
and using a toy model it is shown how it can renormalize singular repulsive interactions.
} %end of abstract
\maketitle
\section{Introduction}
\label{intro}

Chiral Effective Field Theory ($\chi$EFT) has already a long history applied to the $NN$ system. It was first
proposed by Weinberg~\cite{WEINBERG1990288,WEINBERG19913} in the 1990's motivated by the 
success of Chiral Perturbation Theory ($\chi$PT) in the pion-pion~\cite{GASSER1983321,GASSER1983325,GASSER1985465}
and pion-nucleon~\cite{GASSER1988779} sectors, constraining the interactions to those consistent with Chiral
symmetry. Weinberg realized that they have substantial differences with the $NN$ sector since in
the last one the interaction is stronger and reducible diagrams have a nucleon mass enhancement which breaks
the Chiral expansion. For this reason he proposed to use the rules of $\chi$PT to build the potential instead of the scattering amplitude,
and perform a non-perturbative resummation of reducible diagrams using a Schr\"odinger or
Lippmann-Schwinger equation. This is a similar situation, although for different reasons, to the study of an
electron-proton system. Here the electromagnetic interaction is perturbative, but in order to study the 
Hydrogen atom, a non-perturbative resummation of reducible diagrams with the lowest-order one photon exchange 
interaction is needed. In the $NN$ system also bound states appear, as the deuteron or heavier nuclei, and
this non-perturbative resummation is unavoidable, at least in some channels.

Of course, the two systems are essentially different. While the electromagnetic scattering is perturbative
and renormalizable, the strong $NN$ interaction is non-perturbative and non-renormalizable.
Here we will refer as a renormalizable theory to a theory that has a finite number of divergent
amplitudes that can be renormalized by a finite number of renormalization conditions, which is
the usual concept of renormalizability for a fundamental theory.
Perturbation theory and standard renormalization techniques have shown an unprecedented
accuracy~\cite{KinoshitaQED}. 
Due to the success of QED in the 1950's there was a big effort
to build the strong interaction in terms of nucleons and pions. Unfortunately this effort was unsuccessful
and it started a long history for the study of $NN$ interactions~\cite{Machleidt_2001,Machleidt_2017}.

As mentioned before, $\chi$PT has been shown to be very successful constraining the couplings between pions, or
nucleons and pions, to those consistent with Chiral symmetry. For example, the pseudo-scalar $\pi N$ coupling
is not allowed by Chiral symmetry and the pseudo-vector coupling arises naturally. However this brings us to
the problem in the 1950's, the pseudo-vector coupling is non-renormalizable. The way out of
this dilemma is to consider the theory of nucleons and pions, not as a fundamental theory, but as an
Effective Field Theory (EFT) valid only in the low energy regime. 

In the EFT framework one requires renormalizability in a different way. One in principle has an
infinite number of divergent amplitudes, but it also has an infinite number of terms that can absorb the
divergences. Weinberg's assumption~\cite{WEINBERG1979327} is that if one includes all possible terms allowed by symmetry
requirements, then the theory can be renormalized and produces the most general description of the
system for the energy regime where is valid.

In practice one has to deal with a finite number of amplitudes and a finite number of terms, and 
the theory is organized by a power counting rule, which defines the diagrams and the
terms to be included at a certain order in the expansion. The first power counting rule for the $NN$ system, 
introduced by Weinberg~\cite{WEINBERG1990288}, is based on naive dimensional analysis (NDA), which in the case of a perturbative calculation,
fulfills this requirement. In non-perturbative calculations the situation concerning the power
counting to be assigned to $NN$ contact terms 
is not clear and
there has been a lot of discussion about different power counting rules~\cite{KAPLAN1998390,PhysRevC.74.014003,PhysRevC.83.024003,PhysRevC.84.064002,PhysRevC.84.057001,PhysRevC.85.034002,PhysRevC.86.024001,fbs.54.2175,epja.41.341,Machleidt_2010,PhysRevC.88.054002,epja.56.152}.
However, although NDA could
not be the most efficient and it is not clear if it can be inconsistent in non-perturbative calculations~\cite{KAPLAN1996629,PhysRevC.72.054006},
nowadays is probably the most popular.
This manuscript is not intended to discuss the different power counting rules and NDA will be used 
in the evaluation of the finite-range part of the $NN$ potential.

$\chi$EFT at lowest order has contributions from two contact interactions and the one-pion-exchange interaction.
These interactions are singular so that they can not be included in an Schr\"odinger
or Lippmann-Schwinger equation without regularization since they don't vanish for infinite momentum. 
This is a common situation in nuclear physics and the
way to proceed usually is to include a regulator function that kills the interaction at short range or high
momenta. The regulator function depends on a cut-off scale parameter $\Lambda$ which in many applications
in nuclear physics is on the GeV scale. When singular interactions are present a strong dependence on the
cut-off is unavoidable if no further constrains are used. In the EFT program
one uses $\Lambda$ dependent
contact interactions fixing some low energy observable to achieve regulator independence on the
observables.
At higher orders irreducible loop diagrams appear and the contact terms are also used here to absorb
the infinities generated by such diagrams. NDA allows to do it.

There is a big discussion about how this program should be implemented. There are mainly two
different points of view. On one hand one takes $\Lambda$ bigger than the low-energy scales of the EFT ($\sim m_\pi$
in $\chi$EFT) an smaller than the high energy scale of the EFT ($\sim 4\pi f_\pi\sim 1$ GeV in $\chi$EFT).
Then one varies the parameter $\Lambda$ in this window, always fixing the same low energy constraints, and finds
an approximate cut-off independence in a rather small interval of possible values of the cut-off. 
This approach has been shown to be phenomenologically very fruitful and
the high-precision $NN$ potentials in $\chi$EFT are based on this point of view.

On the other hand, one can follow the standard renormalization procedure and take $\Lambda$ to infinity, so
removing completely the cut-off dependence. This is the way QED is renormalized and has been shown to be very
successful in this renormalizable theory. However in $\chi$EFT there is a serious problem, since the
non-perturbative resummation of reducible diagrams can not be performed analytically, and one has to
use numerical methods to sum them. 

Which is the correct approach is something under debate and there have been arguments in favor and against
both approaches. We don't want to enter on this discussion here and refer the reader to some recent review~\cite{10.3389/fphy.2020.00079}.
Nonetheless,  the scattering amplitudes are constrained by unitarity and analyticity which
allows us to introduce a new method~\cite{ENTEM2017498,OLLER2019167965}. The  former relies on a derivation of the exact discontinuity
along the left-hand cut of a two-body scattering amplitude together with the
implementation of unitarity and analyticity by employing the $N/D$ method~\cite{PhysRev.119.467}.

In this work we only want to give an overview of some of the approaches that have been developed
to do the non-perturbative resummation with singular interactions using the second approach. Unfortunately,
these rely on numerical solutions of scattering problem and there is no proof of the exact convergence
of the result in the limit $\Lambda\to\infty$. However, as we will see, quite different approaches agree
for high values of the $\Lambda$ parameter, which gives confidence about the convergence of the result
for the low energy regime. Also a toy model with an exact solution will be employ to check the agreement
of renormalization of singular interactions with the underlying theory.

The paper is organized as follows. 
In Section~\ref{potentials} we write down the $\chi$EFT potentials up to NNLO to be used in all calculations.
In Section~\ref{boundary} we introduce the method of renormalization
with boundary conditions performed in coordinate space. The problem of renormalization of singular repulsive
interactions will be presented using a toy model.
In Section~\ref{onecounter} we turn into momentum
space and show how one can renormalize with one counter term and the equivalence with the previous approach
and introduce substractive renormalization. In Section~\ref{ND} we present the more recent approach based on the $N/D$ method,
but considering the non-perturbative resummation also along the left-hand cut. 
We will show the equivalence with previous approaches
and show how it can go beyond. We will use a toy model to check the agreement of renormalization of singular
interactions with the underlying theory.
We will end in Section~\ref{summary} with a summary
and conclusions.

\section{$\chi$EFT potentials up to NNLO}
\label{potentials}

$\chi$EFT potentials were derived long time ago. The first contributions were given by Weinberg~\cite{WEINBERG1990288,WEINBERG19913}
and very soon after the first nuclear potentials were obtained by Ordo\~nez and van Kolck~\cite{ORDONEZ1992459,PhysRevLett.72.1982,PhysRevC.53.2086}
in coordinate space up to NNLO and regularized by a cut-off function. After that, momentum space potentials
using dimensional regularization for loop diagrams were developed~\cite{KAISER1997758,EPELBAOUM1998107,EPELBAUM2000295} also up to NNLO.
However it was not until 2003 that $\chi$EFT reached high precision when the first chiral potential
at N$^3$LO was developed by the Idaho group~\cite{PhysRevC.68.041001,MACHLEIDT20111} that was able to describe the $NN$ scattering
data with a $\chi^2/{\rm d.o.f}$ of the order of one, similar to what the high-precision potentials of the 90's had achieved~\cite{PhysRevC.49.2950,
PhysRevC.51.38,PhysRevC.53.R1483,PhysRevC.63.024001}. Very soon after the Bochum group developed an N$^3$LO potential~\cite{EPELBAUM2005362} and
during the last years calculations have gone up to N$^4$LO~\cite{PhysRevC.91.014002} and
potentials at this order have been developed~\cite{PhysRevC.96.024004,Reinert2018}.

Since different potentials
used sometimes slightly different power countings, we give here the potentials up to NNLO that are
going to be used in all the present calculations. Here we only give finite range interactions, zero
range interactions given by the contact terms, will be included by renormalization conditions.

We write the potential in momentum space in the usual form
\begin{eqnarray}
V({\vec p}', \vec p) &  = &
 \:\, V_C \:\, + \vec{\tau}_1 \cdot \vec{\tau}_2 \, W_C
\nonumber \\ &+& 
\left[ \, V_S \:\, + \vec{\tau}_1 \cdot \vec{\tau}_2 \, W_S
\,\:\, \right] \, 
\vec\sigma_1 \cdot \vec \sigma_2
\nonumber \\ &+&
\left[ \, V_{LS} + \vec{\tau}_1 \cdot \vec{\tau}_2 \, W_{LS}
\right] \,
\left(-i \vec S \cdot (\vec q \times \vec k) \,\right)
\nonumber \\ &+&
\left[ \, V_T \:\,     + \vec{\tau}_1 \cdot \vec{\tau}_2 \, W_T
\,\:\, \right] \,
\vec \sigma_1 \cdot \vec q \,\, \vec \sigma_2 \cdot \vec q
\nonumber \\ &+&
\left[ \, V_{\sigma L} + \vec{\tau}_1 \cdot \vec{\tau}_2 \,
      W_{\sigma L} \, \right] \,
\vec\sigma_1\cdot(\vec q\times \vec k\,) \,\,
\vec \sigma_2 \cdot(\vec q\times \vec k\,)
\, ,
\end{eqnarray}

At LO there is only the one-pion exchange interaction given by
\begin{eqnarray}
        W_T^{\rm LO} = -\frac{g_A^2}{4f_\pi^2} \frac{1}{q^2 + m_\pi^2} \,,
\end{eqnarray}
where $\vec q$ is the momentum transfer. 

At NLO one-loop contributions with lowest-order $\pi N$ vertexes appear.
The contribution is given by
\begin{eqnarray}
	W_C^{\rm NLO} &=&-{L(q)\over384\pi^2 f_\pi^4} \left[4m_\pi^2(5g_A^4-4g_A^2-1)
+q^2(23g_A^4-10g_A^2-1) + {48g_A^4 m_\pi^4 \over w^2} \right] \,,
\\
V_T^{\rm NLO} &=& -{1\over q^2} V_{S}^{\rm NLO} \; = \; -{3g_A^4 \over 64\pi^2 f_\pi^4} L(q)\,,
\label{eq_2T}
\end{eqnarray}
with
\begin{eqnarray}
        w &=& \sqrt{4m_\pi^2+q^2} \,,
        \\
L(q) &=& {w\over q}
\ln \bigg( \frac{w + q}{2m_\pi} \bigg)
\,,
\end{eqnarray}
At NNLO one-loop diagrams with one vertex of the NLO $\pi N$ Lagrangian appears and give
\begin{eqnarray}
	V_C^{\rm NNLO} &=&  {3g_A^2 \over 16\pi f_\pi^4} 
\bigg[ \frac{g_A^2m_\pi^5}{16m_N w^2} -
\bigg\{ 2m_\pi^2(2c_1-c_3) - q^2 \left( c_3+\frac{3g_A^2}{16m_N}\right) \bigg\}\tilde w^2
A(q) \bigg] \,, 
\\
	W_C^{\rm NNLO} &=&  {g_A^2 \over 128\pi m_N f_\pi^4} 
\bigg[ \frac{3g_A^2m_\pi^5}{w^2} -
	\bigg\{ 4m_\pi^2+2q^2-g_A^2(4m_\pi^2+3q^2) \bigg\}\tilde w^2 \bigg\}
A(q) \bigg] \,, 
\\
V_T^{\rm NNLO} &=&-{1\over q^2}V_{S}^{\rm NNLO} ={9g_A^4 \over 512\pi m_N f_\pi^4} \tilde w^2  A(q)\,,
\\
W_T^{\rm NNLO} &=&-{1\over q^2}W_{S}^{\rm NNLO} =-{g_A^2 \over 32\pi f_\pi^4} 
\bigg[ \bigg( c_4+\frac{1}{4m_N}\bigg)  w^2  -\frac{g_A^2}{8m_N}(10m_\pi^2+3q^2) \bigg] A(q)\,,
\\
V_{LS}^{\rm NNLO} &=&{3g_A^4 \over 32\pi m_N f_\pi^4} \tilde w^2  A(q)\,,
\\
W_{LS}^{\rm NNLO} &=&{g_A^2(1-g_A^2) \over 32\pi m_N f_\pi^4} w^2  A(q)\,,
\end{eqnarray}
with
\begin{eqnarray}
        \tilde w &=& \sqrt{2m_\pi^2+q^2} \,,
        \\
	A(q) &=& {1\over 2q} \arctan\left( \frac{q}{2m_\pi} \right) \,.
\end{eqnarray}

\section{Renormalization with boundary conditions}
\label{boundary}

In this section we will overview the so called renormalization with boundary conditions. This was
already used in molecular physics where singular interactions are usually used. 
The method applied to the $NN$ system was introduce in 
Refs.~\cite{PhysRevC.72.054006,PhysRevC.74.054001,PhysRevC.74.064004}.
In Ref.~\cite{PhysRevC.74.054001} the uncoupled
case is worked out, while the coupled case is covered in Ref.~\cite{PhysRevC.74.064004}.
Here we only
reproduce a few details about the method and some results for comparison with other approaches.
We refer the interested reader to these publications, where many details and
results for all partial waves can be found. 

We will consider for simplicity just the uncoupled case.
It starts with the
Schr\"odinger equation in some partial wave which is written as
\begin{eqnarray}
	\bigg[ -\frac{1}{2m} \frac{d^2}{dr^2}
	+\frac{l(l+1)}{2m} \frac{1}{r^2} + V(r) \bigg] u_{El}(r) = E u_{El}(r)
\end{eqnarray}
When the potential is regular or diverges slower than $1/r^2$ at $r\to0$, near the origin we
have the two well known independent solutions $u_l(r)\sim r^{l+1}$ and $u_l(r)\sim r^{-l}$. The first solution
is regular at the origin while the second gives a non-normalizable wave function and so only
the first one is considered to give physical states. This is the standard way to solve the Schr\"odinger equation.

However when we have a potential that fulfills
\begin{eqnarray}
	\lim_{r\to 0} |r^2 V(r)| = \infty
\end{eqnarray}
the dominant term at the origin is not the centrifugal term any more and the approximate solutions at
the origin are not the previous ones.

Now we have two possibilities. The first one is
\begin{eqnarray}
	\lim_{r\to 0} r^2 V(r) = +\infty
\end{eqnarray}
so we have short range repulsion. In this case the equation can still be solved. There are two solutions, one
convergent to zero and one divergent. Again, the divergent is left out,
and the solution is well defined.

However in the case
\begin{eqnarray}
	\lim_{r\to 0} r^2 V(r) = -\infty
\end{eqnarray}
we have two oscillatory solutions, with oscillations with lower and lower wave lengths when one approaches
the origin. Now the limit of the wave function is not well defined at the origin and the Schr\"odinger
equation can not be solved in the standard way.

The idea of the method is to fix this limit at some energy (usually zero energy) by fixing some low energy
observable (usually scattering length, scattering volume, $\ldots$) and then get the solution at a different
(low)energy using the orthogonality between the wave functions with different energies. This is fulfilled
with the condition
\begin{eqnarray}
	\frac{u'_{E_1l}(r_c)}{u_{E_1l}(r_c)} =
	\frac{u'_{E_2l}(r_c)}{u_{E_2l}(r_c)}
	\label{orthoC}
\end{eqnarray}
at some short-range point $r_c$. Notice that since the Schr\"odinger equation is a linear second order
differential equation the logarithmic derivative at $r_c$ defines uniquely the solution up to the normalization constant.

If we consider the equation at zero energy\footnote{We consider here the case $\lim_{r\to\infty} r V(r)=0$.}
the asymptotic solution for $l=0$ is
\begin{eqnarray}
	\lim_{r\to\infty} u_{00}(r) \to N\bigg( 1-\frac{r}{a} \bigg)
	\label{boundC}
\end{eqnarray}
where $N$ is a normalization constant and $a$ is the scattering length. The method integrates-in the equation
at zero energy with the boundary conditions Eq.~(\ref{boundC}) and using the relation Eq.~(\ref{orthoC}) 
integrates-out the equation to obtain the wave function at a non-zero energy. From the asymptotic 
behavior at $r\to\infty$
the phase-shift can be evaluated.

We first consider the singlet $^1S_0$ partial wave where the LO interaction without contact terms is
regular. The potential is given by
\begin{eqnarray}
	V^{\rm LO}_{1S_0} (r) &=&  -\frac{g_A^2 m_\pi^3}{16 \pi f_\pi^2} \frac{e^{-x}}{x}
	\label{OPE1S0}
\end{eqnarray}
with $x\equiv m_\pi r$.

\begin{figure}
\resizebox{0.75\columnwidth}{!}{
\includegraphics{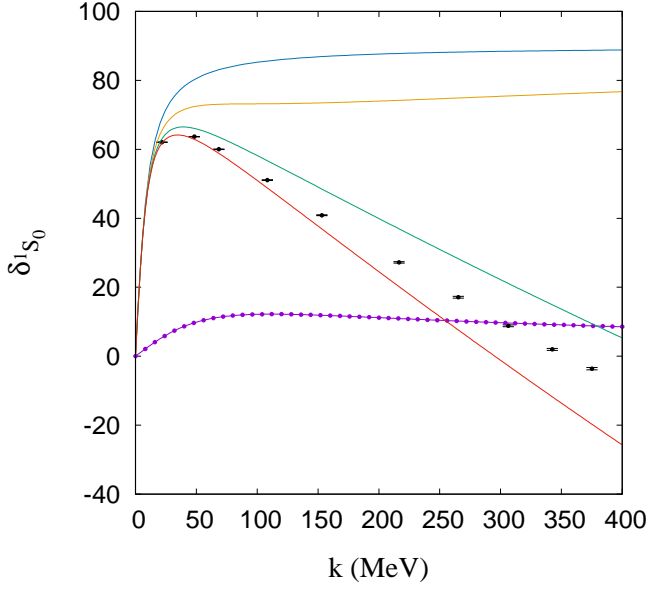} }
\caption{Phase shift in the $^1S_0$ $NN$ partial wave. Black dots with error bands correspond to the results
from the Granada $np$ phase-shift analysis~\cite{PhysRevC.88.064002}. 
The purple dots shows the result for OPE with regular boundary conditions, and the purple line
fitting to $a=-0.91299$ fm.
The blue line corresponds to the effective range expansion at lowest order using the scattering
length from the Granada analysis $a=-23.735$ fm.
The gold, green and red lines give the result for renormalization with boundary condition
of OPE, NLO and NNLO $\chi$EFT potentials, respectively.
}
\label{fig1}       % Give a unique label
\end{figure}

For regular interactions one of course recover the original solution if one fix the scattering length
to the solution of the regular interaction. This is shown in Fig.~\ref{fig1} where we show results for
the potential Eq.~(\ref{OPE1S0}) 
using the parameters of Table~\ref{tab1}.
The purple dots show the solution with standard regular boundary conditions and 
the purple line the result 
fixing the scattering length to the solution of the potential with a value $a=-0.91299$ fm.

\begin{table}
\caption{Parameters used in the calculations.}
\label{tab1}       % Give a unique label
\begin{tabular}{lll}
\hline\noalign{\smallskip}
$\hbar c$ & 197.326963 & MeV$\cdot$fm \\
$g_A$ & 1.26  \\
$m_\pi$ & 138.039 & MeV  \\
$f_\pi$ & 92.4 & MeV  \\
$m_N$ & 938.919 & MeV \\
\noalign{\smallskip}\hline
$c_1$ & -0.74 & GeV$^{-1}$ \\
$c_3$ & -3.61 & GeV$^{-1}$ \\
$c_4$ &  2.44 & GeV$^{-1}$ \\
\noalign{\smallskip}\hline
\end{tabular}
\end{table}

However the method can generate different solutions that will differ by a zero
range interaction in the potential as we will see later. We can fix the scattering length to the Granada value $a=-23.735$ fm
and we get the gold line. 
We include for comparison the result in pionless EFT fixing the same value
of the scattering length with a blue line and the Granada $np$ phase-shift analysis~\cite{PhysRevC.88.064002}
by the solid dots with error bars.

However if we consider the potentials at NLO or NNLO the interactions are singular and attractive so one low energy parameter has to be fixed.
The $r$-space representation of the potentials of Sec.~\ref{potentials} was also given in Ref.~\cite{KAISER1998395}. 
Since
the potentials are divergent in the momentum transfer, the Fourier transform has to be done using a spectral representation
with the necessary subtractions to remove the divergency that only affect zero-range interactions.

We give here the potentials at NLO and NNLO in the $^1S_0$ partial wave for completeness
\begin{eqnarray}
	V^{\rm NLO}_{^1S_0} (r) &=& -\frac{m_\pi^5}{128 \pi^3 f_\pi^4}\frac{1}{x^4}
	\bigg[ (g_A^4(59+4x^2) -10g_A^2-1) x K_0(2x) 
		\nonumber \\ && 
	+ (g_A^4(59+36x^2) -2g_A^2(5+x^2)-1) K_1(2x) \bigg]
	\\
	V^{\rm NNLO}_{^1S_0} (r) &=& \frac{m_\pi^6}{1024 \pi^2 m_N f_\pi^4} \frac{e^{-2x}}{x^6}
	\bigg[ g_A^4 (360 + 720 x + 644 x^2 + 328 x^3 + 96 x^4 + 9 x^5)
		\nonumber \\ &&
		+8 g_A^2 (
		24 c_1 m_N x^2 (x+1)^2 -(12 + 24 x + 20 x^2 + 8 x^3+x^4) 
		\nonumber \\ &&
   - 8 c_4 m_N (3 + 6 x + 5 x^2 + 2 x^3) 
   +12 c_3 m_N (6 + 12 x + 10 x^2 + 4 x^3 + x^4)) 
   \bigg]
		\nonumber \\ &&
\end{eqnarray}
At the origin these potential go as
\begin{eqnarray}
	V^{\rm NLO}_{^1S_0} (r) &\to& -\frac{1}{256 \pi^3 f_\pi^4}\frac{59 g_A^4 -10g_A^2-1}{r^5} =
	-\frac{R_{\rm NLO}^4}{r^5}
%	(59 g_A^4 -10g_A^2-1) 
	\\
	V^{\rm NNLO}_{^1S_0} (r) &\to&
	\frac{3}{128 \pi^2 m_N f_\pi^4} \frac{15g_A^4+4g_A^2(6c_3m_N-2c_4m_N-1)}{r^6}
	=-\frac{R_{\rm NNLO}^5}{r^6}
\end{eqnarray}
with $R_{\rm NLO}=0.767$ fm and $R_{\rm NNLO}=1.056$ fm.

Using these potentials we obtain in Fig.~\ref{fig1} the green line for the NLO case and the red line for the NNLO case. 
The convergence in $r_c\to0$ will be shown in the next section comparing with substractive renormalization.
As mentioned before a careful study of the $NN$ system in the framework of $\chi$EFT up to NNLO was performed
in Refs.~\cite{PhysRevC.74.054001,PhysRevC.74.064004}.

We end this section considering the toy model inspired in the one proposed in Ref.~\cite{Epelbaum_2018}. There a singular
repulsive interactions at long range is regulated by short range interactions which showed a terrible convergence due
to the repulsive character of the interaction. Here we want to show how a repulsive singular interaction can be also renormalized.
For that we consider a long range regular attractive interaction and a two-pion inspired singular repulsion given by the potentials 1 and 2
in Eq.~(\ref{potTM}). The short range interaction 3 and 4 are added to end with a regular interaction at short range so it can be solved and
we will consider as the underlying theory. The expressions are
\begin{eqnarray}
	V(r) &=& V_1(r) +  V_2(r) +  V_3(r) +  V_4(r) 
	\label{potTM}
	\\
	V_1(r) &=& -\alpha \frac{e^{-m_\pi r}}{r}
	\\
	V_2(r) &=& \alpha_1 \frac{e^{-2m_\pi r}}{r^3}
	\\
	V_3(r) &=& -\alpha_1 (m_2-2m_\pi) \frac{e^{-m_1r}}{r^2}
	\\
	V_4(r) &=& -\alpha_1 \frac{e^{-m_2r}}{r^3}
\end{eqnarray}
Once we add $V_2(r)$ the potential is singular repulsive if $V_4(r)$ is not included, however at short 
distances the full potential tends to
\begin{eqnarray}
	V(r) &\to& \frac{1}{2} \frac{-2\alpha +\alpha_1(2m_\pi-m_2)(2m_\pi-2m_1+m_2)}{r}
\end{eqnarray}
We use the parameters of Table~\ref{tab2}.

\begin{table}
\caption{Parameters used in the toy model when applied in the repulsive singular case.}
\label{tab2}       % Give a unique label
\begin{tabular}{lll}
\hline\noalign{\smallskip}
$m_\pi$ & 138.5 & MeV  \\
$m_1$ & 1000 & MeV  \\
$m_1$ & 1200 & MeV  \\
$m_N$ & 938.919 & MeV \\
$\alpha$ & 0.1 & \\
$\alpha_1$ & 5.0 & GeV$^{-2}$ \\ 
\noalign{\smallskip}\hline
\end{tabular}
\end{table}

\begin{figure}
\resizebox{0.75\columnwidth}{!}{
\includegraphics{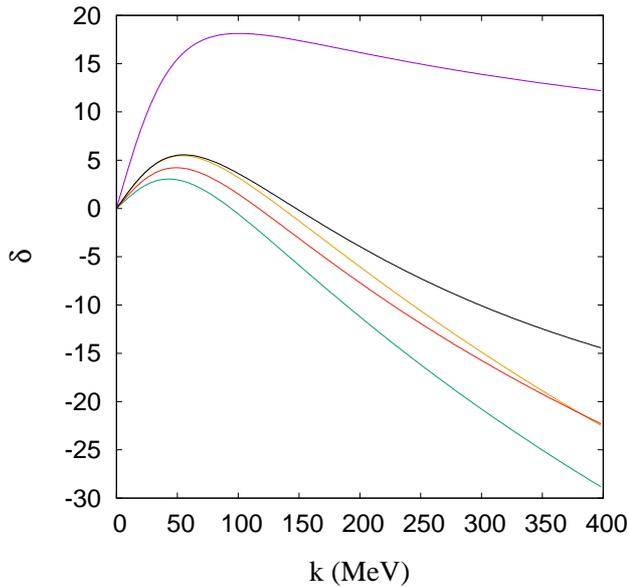} }
\caption{Phase shift for the toy model Eq.~(\ref{potTM}) in $S$ wave.
The purple line shows the regular solution for $V_1(r)$ and the gold line fixing
the scattering length parameter to the solution of the full potential. The green and red lines
show the results summing up to $V_2(r)$ and $V_3(r)$, respectively, while the black line shows the result of the
full theory.
}
\label{fig4}       % Give a unique label
\end{figure}

Although there is no power counting here, in the following we will refer as order $n$ to the result considering 
the sum of the potentials up to $n$ ($\sum_{i=1}^n V_i(r)$).

As order 1 is regular it has a regular solution showed by the purple line in Fig.~\ref{fig4}. The scattering length
is $a=-1.50659$ fm, far from the solution of the full theory $a=-0.615182$ fm. Being regular we can renormalize it
with boundary conditions and we get the gold line which shows the correct low energy behavior, to compare with the
black line which shows the full theory. However orders 2 and 3 are singular repulsive and can not be renormalized
in $r$ space. The results are given by the green line for order 2 and the red line for order 3. Clearly the description
of the system is much worse than the renormalization of the order 1 potential. However we will come back to this issue on Section~\ref{ND}.

Here only scattering states have been considered, however
the method can also be used to study bound states. An example is the application
of the method to study quarkonium models done in Refs.\cite{PhysRevD.85.074001,PhysRevD.86.094027}.

\section{Renormalization with one counter term}
\label{onecounter}

In this section we are going to see how one can performed equivalent calculations in momentum space to 
those in the previous
section. The starting point is the Lippmann-Schwinger equation in momentum space given by
\begin{eqnarray}
	T(p',p;k) &=& V(p',p) +\frac{m_N}{2\pi^2} \int_0^\infty d\tilde p\,V(p',\tilde p) 
	\frac{\tilde p^2}{k^2-\tilde p^2+i\epsilon} T(\tilde p,p;k) 
\end{eqnarray}
where $V(p',p)$ represents the potential projected in one uncoupled partial wave.

The solution of this equation is obtain numerically, only special cases can be performed analytically. In
particular pionless $\chi$EFT is one of such cases. A general procedure to solve these
equation for contact interactions is given in Refs.~\cite{OLLER2020103728,PHILLIPS1998255}.
If one considers
the lowest order in $S$ waves the potential is just a constant $C$ and the equation is
\begin{eqnarray}
	T(p',p;k) &=& C +\frac{m_N}{2\pi^2} \int_0^\infty d\tilde p\,C
	\frac{\tilde p^2}{k^2-\tilde p^2+i\epsilon} T(\tilde p,p;k) 
\end{eqnarray}
To solve the equation one has first to regularize the integral introducing a cutoff scale $\Lambda$, for example
introducing a sharp cutoff in the integration momentum
\begin{eqnarray}
	T_\Lambda(p',p;k) &=& C(\Lambda) +\frac{m_N}{2\pi^2} \int_0^\Lambda d\tilde p\,C(\Lambda)
	\frac{\tilde p^2}{k^2-\tilde p^2+i\epsilon} T_\Lambda(\tilde p,p;k) 
\end{eqnarray}
where the solution of the original equation is obtained in the limit $\Lambda\to\infty$.
The solution is just
\begin{eqnarray}
T_\Lambda(p',p;k) &=& \frac{C(\Lambda)}{1-C(\Lambda) I(k,\Lambda)}
\end{eqnarray}
with
\begin{eqnarray}
I(k,\Lambda) &=&
\frac{m_N}{2\pi^2} \int_0^\Lambda dq
\frac{q^2}{k^2-q^2+i\epsilon}
=
\frac{m_N}{2\pi^2} (-\Lambda +\frac{k}{2} \log \frac{\Lambda+k}{\Lambda-k} )
-i\frac{m_N}{4\pi} k
\label{loopf}
\end{eqnarray}
Since the integral is linearly divergent, the loop function Eq.~(\ref{loopf}) diverges for $\Lambda\to\infty$
and we don't have a meaningful result for a fix value of $C$ in that limit. 
This is the same problem as in usual perturbative
renormalization. The way to solve it is to fix some quantity to its physical value and remove the infinities
with the constants of the Lagrangian. For this reason we already include a $\Lambda$ dependence on the
coupling $C$ which is given by the zero order Lagrangian. Now we fix the $T$ matrix at some scale $\mu$
\begin{eqnarray}
T(\mu) &=&T_\Lambda(p',p;\mu) =
\frac{C(\Lambda,\mu)}{1-C(\Lambda,\mu) I(\mu,\Lambda)}
\end{eqnarray}
which implies
\begin{eqnarray}
C^{-1}(\Lambda,\mu) = T^{-1}(\mu) + I(\mu,\Lambda)
\end{eqnarray}
The solution is now
\begin{eqnarray}
T_\Lambda(p',p;k) &=& 
\frac{1}{T^{-1}(\mu) + I(\mu,\Lambda) - I(k,\Lambda)}
\end{eqnarray}
with
\begin{eqnarray}
I(\mu,\Lambda) - I(k,\Lambda) =
\frac{m_N}{2\pi^2}
(\frac{\mu}{2} \log \frac{\Lambda+\mu}{\Lambda-\mu} -
\frac{k}{2} \log \frac{\Lambda+k}{\Lambda-k} )
-i\frac{m_N}{4\pi} (\mu-k)
\end{eqnarray}
and the result is finite in the limit $\Lambda\to\infty$
\begin{eqnarray}
\lim_{\Lambda\to\infty}
I(\mu,\Lambda) - I(k,\Lambda) &=&
-i\frac{m_N}{4\pi} (\mu-k)
\\
\lim_{\Lambda\to\infty}
T_\Lambda(p',p;k) &=&
\frac{1}{T^{-1}(\mu) -i\frac{m_N}{4\pi} (\mu-k)} =
\frac{T(\mu)}{1 + m_N T(\mu)(ik-i\mu)/4\pi}
\end{eqnarray}
In the case $\mu=0$ one recovers the result given by Weinberg~\cite{WEINBERG19913} and with an imaginary scale
the result of Ref.~\cite{KAPLAN1998329} in dimensional regularization.

This is the procedure of renormalization with a counterterm, one regularizes the Lippmann-Schwinger Equation,
then fixes the value of the contact term to some low energy observable, and finally takes the limit
$\Lambda\to\infty$ to obtain regularization independent results. The problem is that once pions are included
the process can not be performed analytically and only numerical solutions are possible. Also, one can not
use the Fredholm theorem to prove that the solution exists, since the kernel is not square integrable with singular
interactions in that limit. However, there is also no proof that the limit does not exist.

The idea is to consider a potential
\begin{eqnarray}
	V(p',p;\Lambda) &=& V_{\rm ct}(p',p;\Lambda) + V_\pi (p',p;\Lambda) 
	\label{pot}
\end{eqnarray}
where $V_{\rm ct}(p',p;\Lambda)$ are contact terms\footnote{These are zero-range interactions in the $\Lambda\to\infty$
limit} and $V_\pi (p',p;\Lambda)$ finite range interactions
in $\chi$EFT. To be more specific at lowest order in the $^1S_0$ partial wave\footnote{Notice that we have
included contact contributions from OPE in the contact interaction.}
\begin{eqnarray}
	V_{\rm ct}(p',p;\Lambda) &=& C_{^1S_0}(\Lambda) f(p',p;\Lambda) 
	\\
	V_\pi (p',p;\Lambda) &=& -\frac{g_A^2}{4f_\pi^2} \frac{m_\pi^2}{2p'p} Q_0(z) f(p',p;\Lambda) 
	\label{OPEpot}
	\\
	Q_0(z) &=& \frac{1}{2} \bigg( \frac{z+1}{z-1}\bigg)
	\\
	z &=& \frac{p'^2+p^2+m_\pi^2}{2p'p}
\end{eqnarray}
being $f(p',p;\Lambda)$ a regulator function.

This program was performed for the $^1S_0$ $NN$ partial wave in the context of
$\chi$EFT up to N$^3$LO in Ref.~\cite{PhysRevC.77.044006} and for higher partial waves in Ref.~\cite{Zeoli:2012bi}.

A very interesting way to perform this program was proposed in~\cite{FREDERICO1999209}. For a review and the
application of multiple subtractions see~\cite{Batista:2017vao} and references there in. Again here we don't
give all the details of the method and only reproduce those necessary to implement renormalization 
with one counter term
to compare with other methods. The derivations in similar notation as ours can be found 
in~\cite{PhysRevC.77.014002} with slightly
different conventions. We are going to do it also in the simple case of the $^1S_0$ partial wave.

Consider a potential of the type Eq.~(\ref{pot}) with one contact term
\begin{eqnarray}
	V(p',p;\Lambda) &=& ( C + V_\pi(p',p)) \theta(\Lambda-p') \theta(\Lambda-p)
\end{eqnarray}
with $\theta(x)$ the step Heaviside function. Then the Lippmann-Schwinger equation is
\begin{eqnarray}
        T_\Lambda(p',p;k) &=& V_\pi(p',p) + C + \frac{m_N}{2\pi^2} \int_0^\Lambda d\tilde p\,%
        \tilde p^2 \left(\frac{V_\pi(p',\tilde p)+C}{k^2-\tilde p^2+i\epsilon} \right) T_\Lambda(\tilde p,p;k)%
\end{eqnarray}
The method fix $T_\Lambda(0,0;0) = T(0,0;0) =\frac{a}{4\pi m_N}$ and makes the subtraction 
$T_\Lambda(p',0;0)-T_\Lambda(0,0;0)$ to remove $C$ from the equation finding
\begin{eqnarray}
        T_\Lambda(p',0;0) &=& V_\pi(p',0) + \frac{a}{4\pi m_N} +
        \frac{m_N}{2\pi^2} \int_0^\Lambda d\tilde p\,
        \tilde p^2 \left(\frac{V_\pi(p',\tilde p)-V_\pi(0,\tilde p)}{-\tilde p^2} \right) T_\Lambda(\tilde p,0;0)
	\quad 	\quad 
\end{eqnarray}
This equation is solved finding $T_\Lambda(p',0;0)$. Using $T_\Lambda(0,p;0)=T_\Lambda(p,0;0)$ an equation for
the fully off-shell $T$ matrix at zero energy is found
\begin{eqnarray}
        T_\Lambda(p',p;0) - T_\Lambda(0,p;0) &=& V_\pi(p',p)- V_\pi(0,p) 
	\nonumber \\ &&
        + \frac{m_N}{2\pi^2} \int_0^\Lambda d\tilde p\,
        \tilde p^2 \left(\frac{V_\pi(p',\tilde p)-V_\pi(0,\tilde p)}{-\tilde p^2} \right) T_\Lambda(\tilde p,p;0)
\end{eqnarray}
once the $T_\Lambda(p',p;0)$ is found, the $T$ matrix at non-zero energy is obtained solving
\begin{eqnarray}
        T_\Lambda(p',p;k) &=& T_\Lambda(p',p;0) 
	+ \frac{m_N}{2\pi^2} \int_0^\Lambda d\tilde p\,%
	T_\Lambda(p',p;0) \frac{k^2}{k^2-\tilde p^2+i\epsilon} T_\Lambda(\tilde p,p;k)%
\end{eqnarray}
which is similar to a Lippmann-Schwinger equation being $T_\Lambda(p',p;0)$ the potential but with a different
propagator.

\begin{figure}
\resizebox{1.00\columnwidth}{!}{
\includegraphics{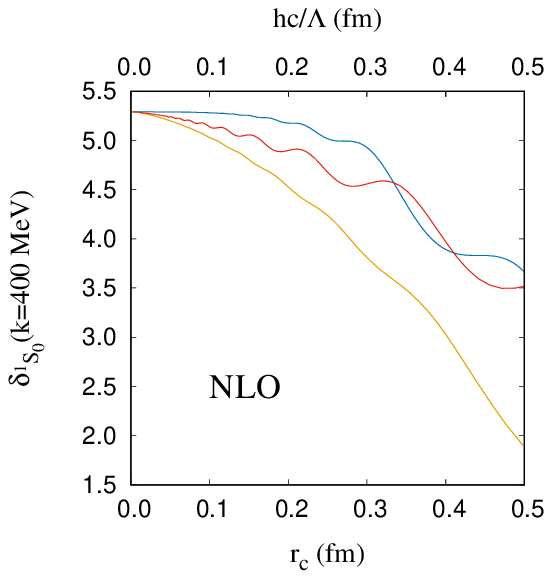} 
\includegraphics{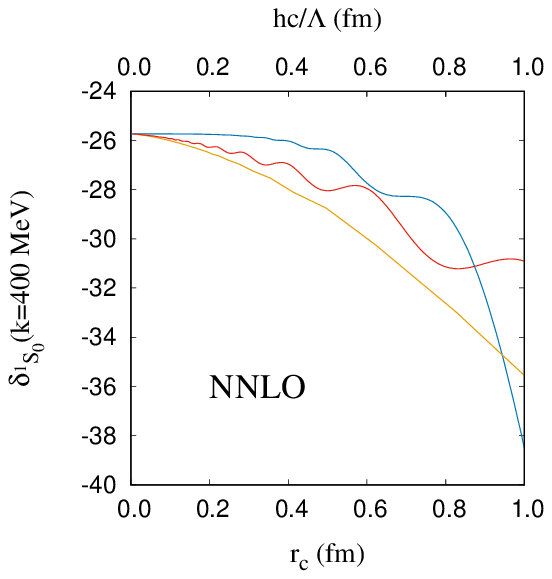} }
\caption{$^1S_0$ $NN$ phase-shifts for the NLO case (left) and NNLO (right) as explained in the text. The blue line
shows the result of renormalization with boundary conditions as a function of the short-range boundary condition point
$r_c$. The gold line shows the results for renormalization with one counter term and a gaussian regulator function Eq.~(\ref{gaussreg})
as a function of the scale $hc/\Lambda$. The red line shows the result of substractive renormalization as explained in the
text.}
\label{fig2}       % Give a unique label
\end{figure}

\begin{figure}
\resizebox{0.75\columnwidth}{!}{
\includegraphics{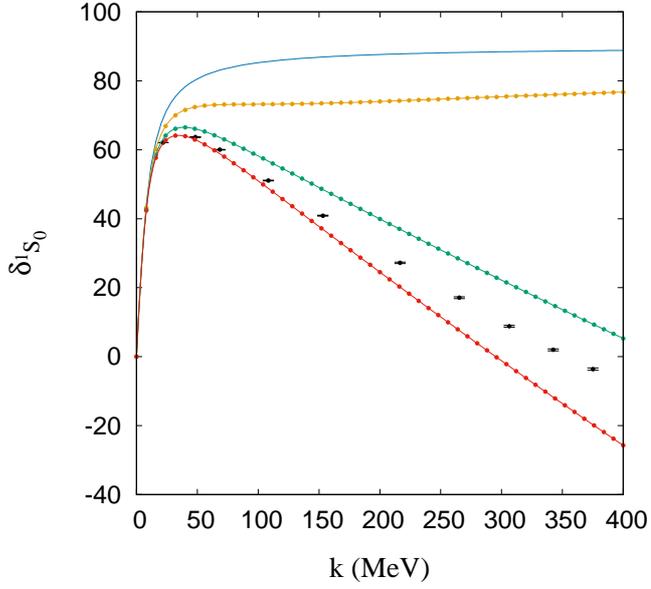} }
\caption{Same color code as in Fig.~\ref{fig1} (gold, green and red for LO, NLO and NNLO respectively)
showing with dots the result of renormalization with boundary conditions and
solid line the result with substractive renormalization.}
\label{fig3}       % Give a unique label
\end{figure}

In Fig.~\ref{fig2} we show results for the NLO and NNLO cases considered previously compared with the
result of renormalization with boundary conditions. We represent the $^1S_0$ phase-shift for onshell
momentum $k=400$ MeV as a function of the boundary condition point $r_c$ and the scale $\frac{hc}{\Lambda}$
for the momentum space calculation. The blue line is the result of the calculation in coordinate space,
the gold line is the result of renormalization with one counter term using 
\begin{eqnarray}
f(p',p;\Lambda)=e^{-\frac{p'^2+p^2}{\Lambda^2}} 
\label{gaussreg}
\end{eqnarray}
and the red line the result of substractive
renormalization. All of them agree in the limit $r_c\to0$ or $\Lambda\to\infty$. In Fig,~\ref{fig3} we compare the
results in coordinate space (dots) with substractive renormalization (solid lines) for LO (gold), NLO (green) and NNLO (red).

In Ref.~\cite{PhysRevC.77.044006} the inclusion of a second contact term to fix the effective range was
investigated and no way was found to fix it with high cutoff values, which is in agreement with the findings in pionless EFT.
 
\section{Non-perturbative calculations with the $N/D$ method}
\label{ND}

In this section we introduce a recent method developed by the authors~\cite{OLLER2019167965} 
to calculate the non-perturbative resummation of reducible
diagrams using the $N/D$ method. 

The $N/D$ method uses the analytical properties of the onshell $T$-matrix. There is always the so called
right-hand-cut (RHC) or physical cut due to intermediate $NN$ states, which starts at the $NN$ threshold
or zero onshell relative $NN$ momentum. However, for interactions in $\chi$EFT there is also a left-hand-cut (LHC)
which is generated by pion exchanges due to the poles of pion propagators. 

The $N/D$ method writes the onshell $T$-matrix in partial waves in the form
\begin{eqnarray}
	T(A) &=& -\frac{N(A)}{D(A)}
\end{eqnarray}
where $A \equiv k^2$ is the square of the onshell momentum and $N(A)$ has only LHC and $D(A)$ RHC.
We have introduced a minus sign since we use a different sign convention for $T(A)$ as it is
usual in $N/D$ method calculations, so $N(A)$ and $D(A)$ are the same as the usual case.

Unitarity on the RHC implies the condition ($A>0$)
\begin{eqnarray}
	{\rm Im} D(A) &=& -\rho(A) N(A)
\end{eqnarray}
with $\rho(A)=\frac{m_N\sqrt A}{4\pi}$ the phase-space factor.

The input of the method is the LHC discontinuity of the $T$-matrix that we denote by $2i\Delta T(A)=2i{\rm Im} T(A)$.
This implies the relation on the LHC ($A<0$)
\begin{eqnarray}
	{\rm Im} N(A) &=& -D(A) \Delta T(A)
\end{eqnarray}

The general expressions of the $N/D$ method with $2n$ subtractions 
were given in Ref.~\cite{PhysRevC.89.014002}.
For regular interactions no subtractions are needed and we can solve 
\begin{eqnarray}
D(A) &=& 1 - \frac{A}{\pi} \int_0^{\infty} d\omega_R \frac{\rho(\omega_R)N(\omega_R)}{(\omega_R-A)\omega_R}
 \\
N(A) &=& -\frac{1}{\pi} \int_{-\infty}^L d\omega_L \frac{D(\omega_L)\Delta(\omega_L)}{(\omega_L-A)}
\end{eqnarray}
Notice that really we made a subtraction in $D(A)$ since $D(A)$ or $N(A)$ has to be fixed at some point.

Now we can make subtractions to fix low energy parameters of the effective range expansion
\begin{eqnarray}
  k\cot \delta = -\frac{1}{a} + \frac 1 2 r k^2 + \sum_{i=2} v_i k^{2i}
  \label{Ef}
\end{eqnarray}
We obtain~\cite{ENTEM2017498} with one subtraction
\begin{eqnarray}
        D(A) &=& 1 + i {a} \sqrt A
  - i\frac{m_N}{4\pi^2} \int_{-\infty}^L d\omega_L
  \frac{D(\omega_L)\Delta (\omega_L)}{\omega_L} \frac{A}{\sqrt A+\sqrt{\omega_L}}
 \\
 N(A) &=& -\frac{4\pi {a}}{m_N}
 - \frac{A}{\pi} \int_{-\infty}^L d\omega_L \frac{D(\omega_L)\Delta(\omega_L)}{(\omega_L-A)\omega_L}
\end{eqnarray}
However we can also obtain equations to fix the effective range $r$, the next effective range parameter $v_2$, etc.
As in Ref.~\cite{ENTEM2017498} we denote by $N/D$$_{01}$\footnote{In general
	$N/D_{mn}$ refers to the $N/D$ dispersion relations with $m(n)$ subtractions in $N(D)$.}
	the regular case, $N/D$$_{11}$ one subtraction to fix $a$,
$N/D$$_{12}$ two subtractions and $N/D$$_{22}$ three subtractions.
Here we will only make up to three subtractions. 

In order to solve the $N/D$ method an integration in an infinite interval for $\omega_L$ is needed. One can use a cut-off there but
we use a mapping of gaussian points in the interval $(-\infty,L)$ and we check convergence with the number of 
gaussian points, so no
cutoff scale is introduced.

In order to solve the $N/D$ method we need to know $\Delta(A)$. To illustrate how the LHC arises let's consider
the onshell OPE potential in the $^1S_0$ partial wave, which is given by
\begin{eqnarray}
	V_\pi^{^1S_0}(A) = -\frac{g_A^2}{4f_\pi^2} \frac{m_\pi^2}{4A} 
\log\bigg(\frac{m_\pi^2+4A}{m_\pi^2} \bigg) 
\end{eqnarray}
where 
again we have not included the contact contribution. When we do the analytical continuation to the 
complex $A$ plane we have the cut due to the $\log$ function. So for
\begin{eqnarray}
	m_\pi^2 + 4A<0 \Rightarrow A<-\frac{m_\pi^2}{4} \equiv L
\end{eqnarray}
the $\log$ cut makes a discontinuity in the imaginary part. The imaginary part on the cut (as the limit from the
upper half plane) is given by
\begin{eqnarray}
	\Delta V_\pi^{^1S_0}(A) = -\frac{\pi g_A^2m_\pi^2}{16f_\pi^2 A} \theta(L-A)
\end{eqnarray}
Notice that here we use a different prescription (with a minus sign difference) 
than in Ref.~\cite{ENTEM2017498,OLLER2019167965} to use
the same prescription for the potential as in previous sections. Also notice that any polynomial in external
momenta generated by contact (counter) terms does not generate any LHC.

In Ref.~\cite{OLLER2019167965} the integral equation to obtain $\Delta T(A)$ in the LHC for the non-perturbative
resummation of reducible diagrams was derived. To see how the method works let's consider the 
onshell once-iterated OPE
in the $^1S_0$ partial wave given by
\begin{eqnarray}
	V^{^1S_0}_{{\rm it}\pi}(k,k) &\equiv&
	\frac{m_N}{2\pi^2} \int _0^\infty dp \, V^{^1S_0}_\pi(k,p) \frac{p^2}{k^2-p^2+i\epsilon'} V^{^1S_0}_\pi(p,k)
	\nonumber \\ &=&
	-\frac{m_N}{4\pi^2} \int _{-\infty}^\infty dp \, 
	V^{^1S_0}_\pi(k,p) \frac{p^2}{p^2-A-i\epsilon'} V^{^1S_0}_\pi(p,k)
\end{eqnarray}
where we have used $V^{^1S_0}_\pi(k,-p)=V^{^1S_0}_\pi(k,p)$ and $k^2=A$. Notice that for $A<0$ there is no pole
for the intermediate $NN$ propagator and we can omit the $-i\epsilon'$ prescription.

To do the analytical continuation we
write the potential as
\begin{eqnarray}
	V^{^1S_0}_{\pi}(p',p) &=& -\frac{g_A^2 m_\pi^2}{16f_\pi^2} \frac{1}{p'p}
	\bigg\{ \log\left[ (p'+p)^2+m_\pi^2 \right] - \log\left[ (p'-p)^2+m_\pi^2 \right] \bigg\}
\end{eqnarray}
Now we consider the onshell momentum imaginary with an small positive real part $k=i\bar k+\epsilon$, so
we approach the LHC from the upper half plane $A=-\bar k^2+i\epsilon$.

Naively one would say that the iterated OPE has no imaginary part since 
\begin{eqnarray}
	V^{^1S_0}_{\pi}(i\bar k,p) &=& -\frac{g_A^2 m_\pi^2}{16f_\pi^2} \frac{1}{i\bar kp}
	\bigg\{ \log\left[ (p+i\bar k)^2+m_\pi^2 \right] - \log\left[ (p-i\bar k)^2+m_\pi^2 \right] \bigg\}
\end{eqnarray}
is real and of course the intermediate $NN$ propagator is also real, so the integral is real. However
this is not the analytical continuation of $V^{^1S_0}_{{\rm it}\pi}(k,k)$ since the integration crosses
the cuts of $V^{^1S_0}_{\pi}(i\bar k,p)$ for $\bar k>m_\pi$. In order to do the analytical continuation
we have to use contour deformations to avoid the cut of the integrand.

\begin{figure}
\resizebox{0.75\columnwidth}{!}{
\includegraphics{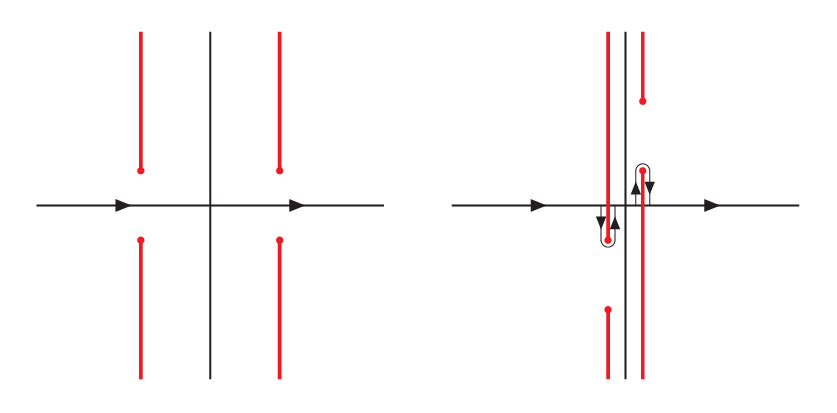} }
\caption{\label{cuts}
Cuts of $V^{^1S_0}_\pi(k,p)$ in the $p$ complex plane for $k$ real (left) and $k=i\bar k+\epsilon$ (right).}
\end{figure}

The branch points of $V^{^1S_0}_{\pi}(k,p)$ are given by $p=-k \pm i m_\pi$ and $p=k\pm i m_\pi$ and the cuts
are represented in Fig.~\ref{cuts} for the case of real $k$ (left) and $k=i\bar k+\epsilon$ (right). There we see
that for real $k$ we don't cross any cuts but for imaginary $k$ we cross them if $\bar k>m_\pi$. So we have to
deform the contour to avoid these cuts as shown in the Figure.

We first perform the integral in the contour with positive real part. There we can use $p=\epsilon-\delta + i\nu$
with $\nu\in(0,\bar k-m_\pi)$ when we go up and $p=\epsilon+\delta + i\nu$ with $\nu\in(\bar k-m_\pi,0)$
when we go down, always with positive $\delta <\epsilon$. The $\log$ functions are then, 
in the limit $\delta \to 0^+$,
\begin{eqnarray}
	\lim_{\delta\to 0^+} \log \big[ (i\bar k + \epsilon +p)^2 +m_\pi^2 \big] &=& 
	\log \big[ m_\pi^2 - (\bar k +\nu)^2 +i\epsilon \big]
	\nonumber \\ &\to &\ln \rho + i\pi
		\\
	\lim_{\delta\to 0^+} \log \big[ (i\bar k + \epsilon -p)^2 +m_\pi^2 \big] &=& 
	\lim_{\delta\to 0^+} \log \big[ m_\pi^2 - (\bar k -\nu)^2 \pm i\delta \big]
	\nonumber \\ &=&
	\ln \big[ (\bar k -\nu)^2 -m_\pi^2 \big] \pm i \pi = \ln \rho' \pm i \pi
\end{eqnarray}
where the $\pm$ sign comes from the path going up $+$ and going down $-$ and 
we define $\rho = (\bar k+\nu)^2-m_\pi^2$
and $\rho' = (\bar k-\nu)^2-m_\pi^2$.
Notice that $\bar k-\nu > m_\pi>0$.

The integrals on the contour are
\begin{eqnarray}
	I_1 &=& \frac{m_N}{4\pi^2} \bigg(\frac{g_A^2m_\pi^2}{16f_\pi^2}\bigg)^2
	\int_0^{\bar k-m_\pi} id\nu\, \frac{\nu^2}{(\bar k+\nu)(\bar k-\nu)} 
	\frac{(\ln \rho-\ln\rho')^2}{\nu^2\bar k^2}
	\\
	I_2 &=& \frac{m_N}{4\pi^2} \bigg(\frac{g_A^2m_\pi^2}{16f_\pi^2}\bigg)^2
	\int^0_{\bar k-m_\pi} id\nu\, \frac{\nu^2}{(\bar k+\nu)(\bar k-\nu)} 
	\frac{(\ln \rho-\ln\rho'+i2\pi)^2}{\nu^2\bar k^2}
	\\
	I_1+I_2 &=&
	\frac{m_N}{\pi} \bigg(\frac{g_A^2m_\pi^2}{16f_\pi^2}\bigg)^2
	\int_0^{\bar k-m_\pi} d\nu\, \frac{1}{(\bar k+\nu)(\bar k-\nu)} 
	\frac{(\ln \rho-\ln\rho')+i \pi}{\bar k^2}
\end{eqnarray}
Doing the same for the other contour we obtain
\begin{eqnarray}
	I_3 &=& -\frac{m_N}{4\pi^2} \bigg(\frac{g_A^2m_\pi^2}{16f_\pi^2}\bigg)^2
	\int_0^{\bar k-m_\pi} id\nu\, \frac{\nu^2}{(\bar k+\nu)(\bar k-\nu)} 
	\frac{(\ln \rho-\ln\rho'+i2\pi)^2}{\nu^2\bar k^2}
	\\
	I_4 &=& -\frac{m_N}{4\pi^2} \bigg(\frac{g_A^2m_\pi^2}{16f_\pi^2}\bigg)^2
	\int^0_{\bar k-m_\pi} id\nu\, \frac{\nu^2}{(\bar k+\nu)(\bar k-\nu)} 
	\frac{(\ln \rho-\ln\rho')^2}{\nu^2\bar k^2}
	\\
	I_3+I_4 &=&
	\frac{m_N}{\pi} \bigg(\frac{g_A^2m_\pi^2}{16f_\pi^2}\bigg)^2
	\int_0^{\bar k-m_\pi} d\nu\, \frac{1}{(\bar k+\nu)(\bar k-\nu)} 
	\frac{(\ln \rho-\ln\rho')+i\pi}{\bar k^2}
\end{eqnarray}
The imaginary part of the iterated OPE is then~\footnote{Again notice that the integral on the real $p$ axis is real.}
\begin{eqnarray}
	{\rm Im}\big[ V^{^1S_0}_{{\rm it}\pi} (A) \big] &=&
	{\rm Im}\big[ I_1+I_2+I_3+I_4 \big] 
	\nonumber \\ &=&
	\frac{2m_N}{\bar k^2} \bigg(\frac{g_A^2m_\pi^2}{16f_\pi^2}\bigg)^2
	\int_0^{\bar k-m_\pi} d\nu\, \frac{1}{(\bar k+\nu)(\bar k-\nu)} 
	\nonumber \\ &=&
	\frac{2m_N}{\bar k^2} \bigg(\frac{g_A^2m_\pi^2}{16f_\pi^2}\bigg)^2
	\frac{1}{2\bar k} \log\bigg( \frac{2\bar k}{m_\pi} - 1 \bigg)
	\nonumber \\ &=&
	-\frac{m_N}{A\sqrt{-A}} \bigg(\frac{g_A^2m_\pi^2}{16f_\pi^2}\bigg)^2
	\log\bigg( \frac{2\sqrt{-A}}{m_\pi} - 1 \bigg) \theta (4L-A)
\end{eqnarray}
where at the end we include the Heaviside step function to make explicit the condition $\bar k > m_\pi$.

The integral equation to obtained $\Delta T(A)$ was deduced in~\cite{OLLER2019167965} using similar techniques
and the properties of the half-offshell $T$-matrix. We write here Eq.~(5.47) for the uncoupled 
case\footnote{We use a different notation than in~\cite{OLLER2019167965}, so is more in tune with previous sections}
\begin{eqnarray}
	\Delta \hat T(\nu,\bar k) &=& \Delta \hat V(\nu,\bar k) - \theta(\bar k-m_\pi) 
	\theta(\bar k-2m_\pi-\nu) \frac{m_N}{2\pi^2} 
	\nonumber \\ &&
	\int_{\nu+m_\pi}^{\bar k-m_\pi} d\nu_1 \,\frac{\nu_1^2}{\bar k^2-\nu_1^2} S(\nu_1)
	\Delta \hat V(\nu,\nu_1) \Delta \hat T(\nu_1,\bar k)
	\label{IntEq}
\end{eqnarray}
with
\begin{eqnarray}
	S(\nu_1) &=& \frac{1}{(\nu_1+i\epsilon)^{2l+2}} + \frac{1}{(\nu_1-i\epsilon)^{2l+2}} 
	\\
	2\Delta \hat V(\nu_1,\nu_2) &=& \nu_1^{l+1} \nu_2^{l+1} 
	\lim_{\epsilon\to 0} \lim_{\delta\to 0} 
	\big[
	{\rm Im} V(i\nu_1+\epsilon-\delta,i\nu_2+\epsilon)
	\nonumber \\ && \quad \quad \quad \quad  \quad \quad \quad 
	-{\rm Im} V(i\nu_1+\epsilon+\delta,i\nu_2+\epsilon)
	\big]
	\\
	2\Delta \hat T(\nu,\bar k) &=& \nu^{l+1} \bar k^{l+1} 
	\lim_{\epsilon\to 0} \lim_{\delta\to 0} 
	\big[
	{\rm Im} T(i\nu+\epsilon-\delta,i\bar k+\epsilon)
	\nonumber \\ && \quad \quad \quad \quad  \quad \quad \quad 
	-{\rm Im} T(i\nu+\epsilon+\delta,i\bar k+\epsilon)
	\big]
\end{eqnarray}
where the onshell $T$-matrix is given by
\begin{eqnarray}
	\Delta T(A) &=& -\frac{\Delta \hat T(-\bar k,\bar k)}{\bar k^{2l+2}}
\end{eqnarray}
Notice that the integral equation Eq.~(\ref{IntEq}) is always finite and no cutoff scale is introduced.

If we want to get the once-iterated OPE we only need to calculate the integral term in Eq.~(\ref{IntEq}) changing
$\Delta \hat T(\nu_1,\bar k)$ by $\Delta \hat V(\nu_1,\bar k)$. This is done in appendix~\ref{Ap1}. Also
from Eq.~(\ref{IntEq}) one can demonstrate the Eqs.~(8-9) from~\cite{ENTEM2017498} since it is easy to show
that Eq.~(9) is just Eq.~(\ref{IntEq}) for the case of OPE in the $^1S_0$ partial wave.

The method can be used for any interaction that has a spectral decomposition since the structure of the LHC is
the same. If we consider the spectral decomposition
\begin{eqnarray}
	V(q) &=& \int_{\mu_0}^\infty d\mu\, \frac{2\mu \eta(\mu^2)}{q^2+\mu^2}
\end{eqnarray}
for the $S$ waves we have
\begin{eqnarray}
	\Delta \hat V(\nu_1,\nu_2) &=& -{\rm sign}(\nu_1-\nu_2) \theta(|\nu_1-\nu_2|-\mu_0) \frac{\pi}{2}
	\int_{\mu_0}^{|\nu_1-\nu_2|} d\mu \,\mu\eta(\mu^2)
\end{eqnarray}
For the $^1S_0$ partial wave we get for the NLO contribution
\begin{eqnarray}
	\Delta \hat V_{\rm NLO}^{^1S_0} (\nu_1,\nu_2) &=& 
	-\theta(|\nu_1-\nu_2|-2m_\pi) {\rm sign}(\nu_1-\nu_2)
  \frac{1}{1536\pi f_\pi^4} \frac{\nu_{12}}{\hat \omega(\nu_{12})}
  \bigg\{
   \nonumber \\ &&
          (10+52g_A^2-158g_A^4) m_\pi^4 
%  \nonumber \\ &&
          +\frac 1 2(197g_A^4-46g_A^2-7)  m_\pi^2\nu_{12}^2 
   \nonumber \\ &&
          -\frac 1 4 (59g_A^4-10g_A^2-1)  \nu_{12}^4
%  \nonumber \\ &&
          +6(5g_A^4+2g_A^2+1) 
          m_\pi^4\hat L(\nu_{12}) 
  \bigg\} 
\end{eqnarray}
with $\nu_{12} = |\nu_1-\nu_2|$ and
\begin{eqnarray}
        \hat \omega (\nu_{12}) &\equiv& \sqrt{\nu_{12}^2-4m_\pi^2}
        \\
        \hat L(\nu_{12}) &\equiv& \frac{\hat \omega(\nu_{12})}{\nu_{12}}
        \log\bigg(\frac{\nu_{12}+\hat \omega(\nu_{12})}{2m_\pi}\bigg)
\end{eqnarray}
And for the NNLO contribution
\begin{eqnarray}
	\Delta \hat V^{^1S_0}_{\rm NNLO} (\nu_1,\nu_2) &=&
  -\theta(|\nu_1-\nu_2|-2m_\pi) {\rm sign}(\nu_1-\nu_2)
  \frac{g_A^2}{3840 f_\pi^4}
  \bigg\{
  \nonumber \\ &&
          \bigg( 480 c_1 - 336 c_3 - 128 c_4 - \frac{4}{M_N} - \frac{25g_A^2}{4M_N}\bigg)  m_\pi^5
   \nonumber \\ &&
         +\bigg(-720 c_1 + 360 c_3          - \frac{30}{M_N} + \frac{30g_A^2}{M_N}\bigg)  m_\pi^4\nu_{12}
   \nonumber \\ &&
         +\bigg( 120 c_1 - 120 c_3 + 40 c_4 + \frac{20}{M_N} - \frac{95g_A^2}{2M_N}\bigg)  m_\pi^2\nu_{12}^3
   \nonumber \\ &&
         +\bigg(            18 c_3 -  6 c_4 - \frac{3}{M_N} + \frac{45g_A^2}{4M_N}\bigg)  \nu_{12}^5
  \bigg\}
\end{eqnarray}

\begin{figure}
\resizebox{0.75\columnwidth}{!}{
\includegraphics{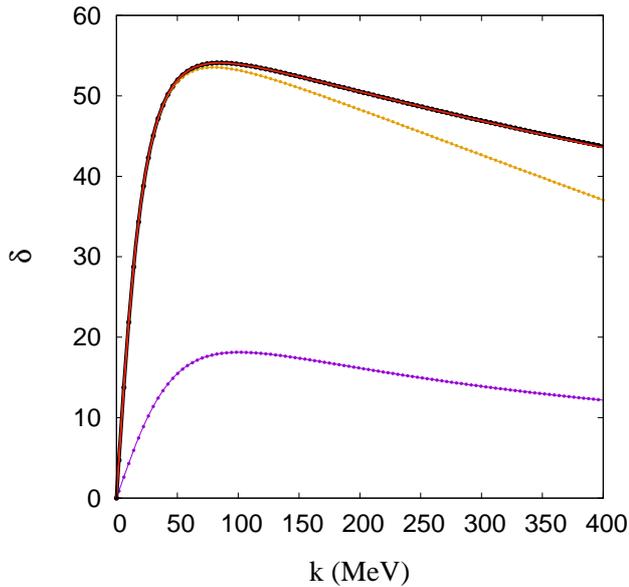} 
}
\caption{Phase shifts for the toy model as in Eq.~(\ref{potTM}) but in the singular attractive case using $\alpha_1=-3$ GeV$^{-2}$. 
The black (purple) line and dots show the result of the full (regular $V_1(r)$) 
potential for the $N/D$ and $r$ space calculations, respectively.
The gold line 
and dots are the same for the case of considering the singular attractive interaction at order 2 with one renormalization
condition. The red line gives the result for the $N/D$$_{22}$ calculation with three renormalization
conditions. The low energy constants are those of the
full theory.
}
\label{fig4b}       % Give a unique label
\end{figure}

For the $^1S_0$ $NN$ partial wave the potentials in $\chi$EFT are singularly attractive. They can be renormalized with one subtraction,
however as shown in Ref.~\cite{ENTEM2017498} additional solutions with more subtractions can be found. Here we consider the toy model
proposed first in Eq.~(\ref{potTM}) but in a singular attractive case by using $\alpha_1=-3$ GeV$^{-2}$. Now the situation is the opposite to the previous case,
order 1 is regular but order 2 is singular attractive. In Fig.~\ref{fig4b} we show the result for the order 2 case. Again the full result is given by
black dots for the $r$ space calculation and a black solid line for the $N/D$ method. 
Similarly the magenta line and dots refer to the results of the regular potential $V_1(r)$ in
$r$ space and with the $N/D$ method, in this order.
The gold line and dots give the result of the
$N/D$ method and boundary conditions, respectively fixing $a$ and considering the order 2 case. $N/D$ allows to make more subtractions
and we perform a calculation with two and three subtractions. The case $N/D$$_{12}$ does not show converge and we don't include it
in the Figure.
The red line give the result with three subtractions $N/D$$_{22}$, fixed in terms of $a$, $r$ and $v_2$,
which agrees very well with the full theory. 
The low-energy parameters in the effective range expansion are taken
from the result of the full theory.

\begin{figure}
\resizebox{0.75\columnwidth}{!}{
\includegraphics{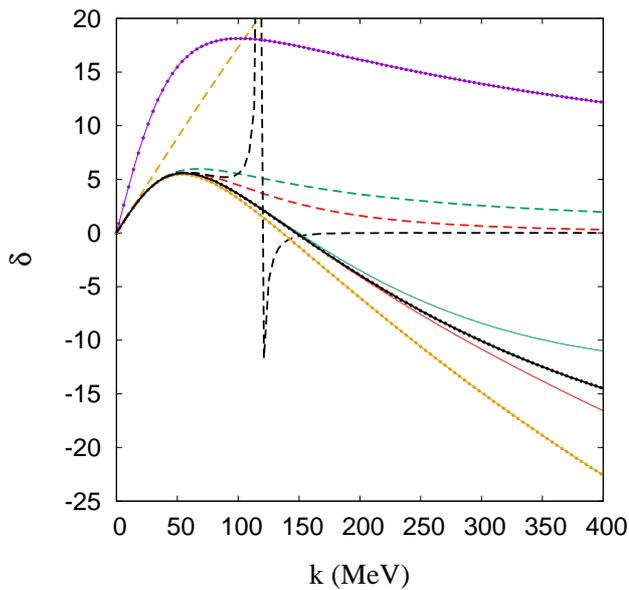} 
}
\caption{Phase shifts for the toy model as in Eq.~(\ref{potTM}). 
The dots shows the result in coordinate space, the full lines are the result of $N/D$ method, and the
	dashed lines the effective range expansion up to $k^{2n}$. The dashed gold, green, red and black are the results for 
	of the effective range expansion for $n=0,1,2,6$,
	respectively. For the coordinate space we have in purple and black the regular solution for $V_1(r)$ and $V(r)$, respectively, and
	in gold fixing $a$ and using only $V_1(r)$. For the $N/D$ method we have in a black line the solution of the full theory, and
	in gold, green and red the solution using $V_1(r)$ with one, two and three subtractions respectively.
}
\label{fig5}       % Give a unique label
\end{figure}

One of the main problems of the previous methods is that they can not renormalize the singular repulsive case. Only the regular
solution is possible, as shown in Fig.~\ref{fig4} for our toy model, and the results including the singular repulsive
interactions does not even have the correct
low-energy behavior. However the exact $N/D$ method allows the renormalization of such cases as we discuss
next.

Let's start showing results for the toy model proposed in Eq.~(\ref{potTM}).
In Figure~\ref{fig5} we show the results for the order 1 case. Again the black line is the full result. The dots correspond to the results
in $r$ space, while the lines shows the results for the regular case, one, two and three subtractions in purple, gold, green and
red lines, respectively. Since order 1 is regular any number of subtractions can be made. We also include for comparison with dashed lines
the result of the effective range expansion with only $a$, adding $r$, adding $v_2$ and up to $v_6$ in gold, green, red and black lines. 
Notice that the parameters of the effective range expansion are not fitted to phase shifts, they are the exact values of the full theory 
given in Tab.~\ref{tab3}.
We can see
that the effective range expansion is only valid for very low energies, however the renormalized results at order 1 give a good
description at higher energies, although for $k\sim 250$ MeV sizable discrepancies with the full theory are seen.
The reason why the effective range expansion is not good is clear, at around $k\sim 150$ MeV, $k\cot\delta$ becomes singular
and the expansion is not valid any more as can be seen from the dashed-black line that goes up to $k^{12}$ in the expansion.
Only considering the finite range interactions, we can go beyond this point.

Another interesting point that can be seen on Fig.~\ref{fig5} is that adding subtraction allows to go to higher energies. The green line
with two subtractions start to show differences with the full theory at the scale of the figure for $k\sim 150-200$ MeV, while the red line
can go up to $k\sim 200-250$ MeV. This is the same idea that was illustrated in Fig.~5 of~\cite{ENTEM2017498}.

\begin{figure}
\resizebox{0.75\columnwidth}{!}{
\includegraphics{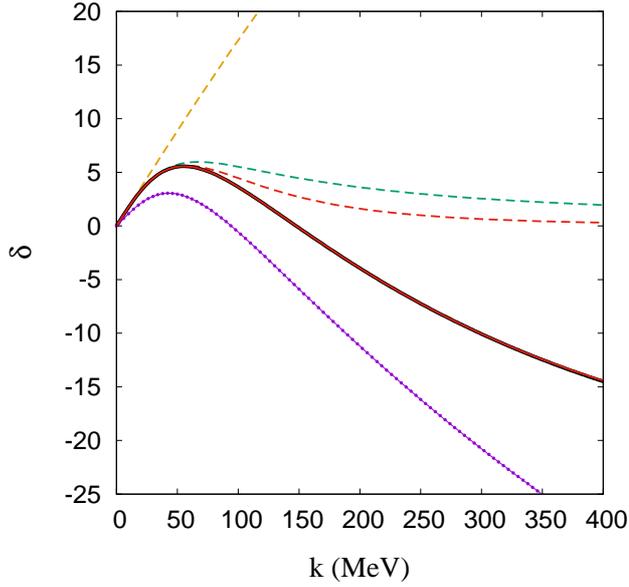} 
}
\caption{Same as in Fig.~\ref{fig5}, but for the order 2 case. The red line is the result of the $N/D$ method with a singular repulsive
interaction $V_1(r)+V_2(r)$ with three subtractions. The low energy constants are not fitted to the phase shifts, they are the
exact low energy constants of the full theory given in Table~\ref{tab3}.
}
\label{fig5b}       % Give a unique label
\end{figure}

However now we consider the order 2 case which adds a singular repulsive interaction. 
The results are given in Fig.~\ref{fig5b} with the same color codes as in Fig.~\ref{fig5}. Being a singular repulsive case, the 
regular solution is the one obtained in coordinate space as can be seen comparing the dots and line in purple. Also we can not fix $a$
as in the previous renormalization methods and so the gold line does not appear. However with the $N/D$ method we can use additional subtractions.
The calculation shows that we don't have convergence fixing $a$ and $r$ and the green line is absent. The result fixing up to $v_2$ converges
and is shown with the red line, which shows a very good agreement with the full theory (black line).

\begin{figure}
\resizebox{0.75\columnwidth}{!}{
\includegraphics{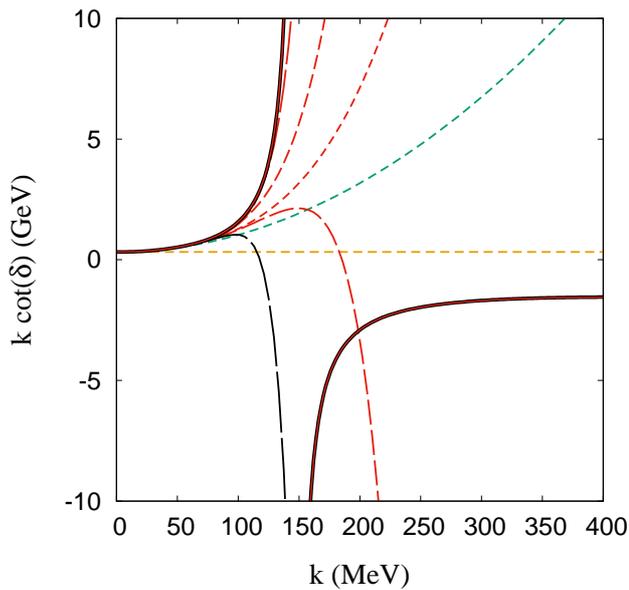} 
}
\caption{$k\cot(\delta)$ effective range expansion for the toy model. The black solid line is the result of the full theory. The dashed lines shows the result
	of Eq.~(\ref{Ef}) cutting the sum at an order $k^{2n}$. The gold, green and red short dashed lines corresponds to $n=0,1,2$ respectively.
	Then the cases $n=3,4,5$ are shown as red dashed lines with longer dashed lengths. The last case shown is $n=6$ in a black dashed line.
	The red solid line shows the result of $N/D_{22}$ at order 2.
}
\label{fig7}       % Give a unique label
\end{figure}

In Table~\ref{tab3} we give the effective range expansion parameters for the different cases. They give a feeling of the accuracy
of the calculation at low energies. Finally in Fig.~\ref{fig7} we show instead the phase-shift, $k\cot(\delta)$
to show the breaking of the effective range expansion in the toy model. 
The black solid line is the result of the full theory. The dashed lines shows the result
of Eq.~(\ref{Ef}) cutting the sum at an order $k^{2n}$. The gold, green and red short dashed lines corresponds to $n=0,1,2$ respectively.
Then the cases $n=3,4,5$ are shown as red dashed lines with longer dashed lengths. The last case shown is $n=6$ in a black dashed line.
The red solid line corresponds to the result renormalizing with three-subtractions the singular repulsive
interaction at order 2.

\begin{table}
	\caption{Effective range parameters as defined in Eq.~(\ref{Ef}) for the toy model Eq.~(\ref{potTM}).}
\label{tab3}       % Give a unique label
\begin{tabular}{llllllll}
\hline\noalign{\smallskip}
& $a$ (fm) & $r$ (fm) & $v_2$ (fm$^3$) & $v_3$ (fm$^5$) & $v_4$ (fm$^7$) & $v_5$ (fm$^9$) & $v_6$ (fm$^{11}$) \\
\hline\noalign{\smallskip}
$N/D$$_{01}$        &&&&&&&\\
 order 1              & -1.50659 & 8.42932 & -5.62505 & 22.3554 & -113.278 & 647.596 &     -3979.14 \\
 order 2              & -0.407820 & 55.2322 & 112.088 & 532.816 & 2114.65 & 10562.8 &     38255.68 \\
\hline\noalign{\smallskip}
$N/D$$_{11}$ &&&&&&&\\
 order 1              & -0.615182 & 28.6632 & 22.3775 & 79.7037 & -4.42830 & 984.991 &     -4097.91 \\
\hline\noalign{\smallskip}
$N/D$$_{12}$ &&&&&&&\\
 order 1              & -0.615182 & 28.1482 & 18.9220 & 63.7131 & -52.0605 & 859.823 &     -4412.90 \\
\hline\noalign{\smallskip}
$N/D$$_{22}$ &&&&&&&\\
 order 1              & -0.615182 & 28.1482 & 19.0066 & 64.7974 & -47.5274 & 871.750 &     -4384.37 \\
 order 2              & -0.615182 & 28.1482 & 19.0066 & 64.7579 & -47.9885 & 870.161 &     -4388.18 \\
\noalign{\smallskip}\hline
Full & -0.615182 & 28.1482 & 19.0066 & 64.7583 & -47.9860 & 870.170 &     -4388.16 \\
\noalign{\smallskip}\hline
\end{tabular}
\end{table}

Now the open question is to know a priory when a solution exists or not in terms of the number of subtractions,
which is indeed an interesting problem in mathematical physics.
It is true that here we don't proof the convergence of the solution, we just rely on numerical convergence, but
the agreement with the full theory is compeling.

\section{Summary and conclusion}
\label{summary}

In this work we have made a brief review of some of the methods used to remove the cut-off dependence when singular
interactions are used. The methods are renormalization with boundary conditions, renormalization in momentum
space with one counter term, and equivalently substractive renormalization, and the exact
$N/D$ method with subtractions. We showed that the methods are equivalent for singular repulsive potentials,
or singular attractive potentials with a renormalization condition.
However, the $N/D$ method allows to make more subtractions or, analogously, implement more 
renormalization conditions.
In particular we made a toy model with
the property that removing
the short range part of the potential is singular, but the full model is regular and can be solved using standard
techniques. We study the singular attractive and repulsive cases, showing that the exact $N/D$ method
with multiple subtractions can describe the full theory in the low energy regime. The convergence of the
solutions are only checked numerically and it is still open to strictly demonstrate that these solutions
exist. However, the agreement with the full theory is compeling, giving hope to renormalize the
$NN$ interaction in the framework of $\chi$EFT.

\begin{acknowledgement}
DRE wants to thank E. Ru\'\i z-Arriola for fruitful discussions about the renormalization with boundary conditions and
with one counter term.
This work has been funded by 
Ministerio de Ciencia e Innovaci\'on
under Contract No. PID2019-105439GB-C22/AEI/10.13039/501100011033,
and PID2019-106080GB-C22/AEI/10.13039/501100011033,
and by EU Horizon 2020 research and innovation program, STRONG-2020 project, under grant agreement No 824093.
\end{acknowledgement}

\appendix

\section{Iterated OPE}
\label{Ap1}

Here we obtain the once iterated OPE from Eq.~(\ref{IntEq}) as
\begin{eqnarray}
	\Delta \hat V^{^1S_0}_{{\rm it}\pi}(-\bar k,\bar k) &=& -\theta(\bar k-m_\pi) 
	\frac{m_N}{2\pi^2} 
	\int_{-\bar k+m_\pi}^{\bar k-m_\pi} d\nu_1 \,\frac{\nu_1^2}{\bar k^2-\nu_1^2} S(\nu_1)
	\nonumber \\ &&
	\Delta \hat V^{^1S_0}_\pi(-\bar k,\nu_1) \Delta \hat V^{^1S_0}_\pi(\nu_1,\bar k)
\end{eqnarray}
We start calculating
\begin{eqnarray}
	2\Delta \hat V^{^1S_0}_\pi (\nu_1,\nu_2) &=& \nu_1 \nu_2
	\lim_{\epsilon\to 0} \lim_{\delta\to 0} 
	\big[
		{\rm Im} V^{^1S_0}_\pi (i\nu_1+\epsilon-\delta,i\nu_2+\epsilon)
	\nonumber \\ && \quad \quad \quad \quad  \quad \quad \quad 
	-{\rm Im} V^{^1S_0}_\pi (i\nu_1+\epsilon+\delta,i\nu_2+\epsilon)
	\big]
\end{eqnarray}
We have that
\begin{eqnarray}
	V^{^1S_0}_\pi (i\nu_1+\epsilon\mp \delta,i\nu_2+\epsilon) &=& -\frac{g_A^2 m_\pi^2}{16 f_\pi^2} 
	\frac{1}{(i\nu_1+\epsilon-\delta)(i\nu_2+\epsilon)}
	\nonumber \\ &&
	\bigg\{
	\log \big[ (i\nu_1+\epsilon\mp \delta+i\nu_2+\epsilon)^2 + m_\pi^2 \big]
	\nonumber \\ &&
	-\log \big[ (i\nu_1+\epsilon\mp \delta-i\nu_2-\epsilon)^2 + m_\pi^2 \big]
	\bigg\}
	\nonumber \\ &\sim&
	\frac{g_A^2 m_\pi^2}{16 f_\pi^2} 
	\frac{1}{\nu_1\nu_2}
	\nonumber \\ &&
	\bigg\{
		\log \big[ m_\pi^2-(\nu_1+\nu_2)^2 +i{\rm sing}(\nu_1+\nu_2)\epsilon\big]
	\nonumber \\ &&
	-\log \big[ m_\pi^2-(\nu_1-\nu_2)^2\mp i{\rm sing}(\nu_1-\nu_2)\delta\big]
	\bigg\}
\end{eqnarray}
and so
\begin{eqnarray}
	\Delta \hat V^{^1S_0}_\pi (\nu_1,\nu_2) &=& 
	{\rm sing}(\nu_1-\nu_2) \theta(|\nu_1-\nu_2|-m_\pi)
	\frac{\pi g_A^2 m_\pi^2}{16 f_\pi^2} 
\end{eqnarray}
From the limits we have
\begin{eqnarray}
	\nu_1 > -\bar k + m_\pi &\Rightarrow& \nu_1+\bar k > m_\pi > 0
	\\
	\nu_1 < \bar k - m_\pi &\Rightarrow& \bar k -\nu_1 > m_\pi > 0
\end{eqnarray}
So we have
\begin{eqnarray}
	\Delta \hat V^{^1S_0}_{{\rm it}\pi}(-\bar k,\bar k) &=& -\theta(\bar k-m_\pi) 
	\frac{m_N}{\pi^2} 
	\int_{-\bar k+m_\pi}^{\bar k-m_\pi} d\nu_1 \,\frac{1}{\bar k^2-\nu_1^2} 
	\bigg( \frac{\pi g_A^2 m_\pi^2}{16 f_\pi^2} \bigg)^2
	\nonumber \\
	&=& -\theta(\bar k-m_\pi) 
	\frac{m_N}{\bar k} 
	\bigg( \frac{g_A^2 m_\pi^2}{16 f_\pi^2} \bigg)^2
	\log \bigg(\frac{2\bar k}{m_\pi}-1\bigg) 
\end{eqnarray}
And finally
\begin{eqnarray}
	\Delta V^{^1S_0}_{{\rm it}\pi}(A) =
	-\frac{\Delta \hat V^{^1S_0}_{{\rm it}\pi}(-\bar k,\bar k)}{\bar k^2}
	&=& \theta(\bar k-m_\pi) 
	\frac{m_N}{\bar k^3} 
	\bigg( \frac{g_A^2 m_\pi^2}{16 f_\pi^2} \bigg)^2
	\log \bigg(\frac{2\bar k}{m_\pi}-1\bigg) 
\end{eqnarray}

\bibliographystyle{apsrev4-1}
\bibliography{Entem}

%\begin{thebibliography}{}
%\bibitem{PhysRevC.74.054001}
%	M. Pav\'on-Valderrama and E. Ruiz-Arriola, Phys. Rev. C \textbf{74}, (2006) 054001:1-39.

%\bibitem{PhysRevC.74.064004}
%	M. Pav\'on-Valderrama and E. Ruiz-Arriola, Phys. Rev. C \textbf{74}, (2006) 064004:1-19.

%Author, Journal \textbf{Volume}, (year) page numbers
%Author, \textit{Book title} (Publisher, place year) page numbers
% etc
%\end{thebibliography}

\end{document}